# High-Throughput Production of Cheap Mineral-Based 2D Electrocatalysts for High-Current-Density Hydrogen Evolution

Chi Zhang[1†], Yuting Luo[1†], Junyang Tan[1], Fengning Yang[1], Zhiyuan Zhang[1], Liusi Yang[1], Hui-Ming Cheng[1,2] & Bilu Liu[1*]

1. Shenzhen Geim Graphene Center, Tsinghua-Berkeley Shenzhen Institute & Tsinghua Shenzhen International Graduate School, Tsinghua University, Shenzhen 518055, P. R. China.

2. Shenyang National Laboratory for Materials Sciences, Institute of Metal Research, Chinese Academy of Sciences, Shenyang, Liaoning, 110016, P. R. China.

† These two authors contributed equally.

* E-mail: bilu.liu@sz.tsinghua.edu.cn

**Abstract**

The high-throughput scalable production of cheap, efficient and durable electrocatalysts that work well at high current densities demanded by industry is a great challenge for the large-scale implementation of electrochemical technologies. Here we report the production of a two-dimensional molybdenum disulfide ($MoS_2$)-based ink-type electrocatalyst by a scalable top-down exfoliation technique followed by a simple heat treatment. The catalyst shows a high current density of 1000 mA cm$^{-2}$ at an overpotential of 454 mV for the hydrogen evolution reaction (HER) without the need of iR compensation, as well as good stability over 24 h. Using the same method, we have, for the first time, produced a cheap $MoS_2$ mineral-based catalyst and found that it had an excellent performance for high-current-density HER. Noteworthy, production rate of this $MoS_2$-based catalyst is one to two orders of magnitude higher than those previously reported. In addition, the price of the $MoS_2$ mineral is five orders of magnitude lower than commercial Pt catalysts, making the $MoS_2$ mineral-based catalyst cheap, and the ink-type catalyst dispersions can be easily integrated with other technologies for large-scale catalyst electrode preparation. These advantages indicate the huge potentials of this method and mineral-based cheap and abundant natural resources as catalysts in the electrochemical industry.

**Introduction**

The large-scale production of hydrogen by electrochemical water splitting has been proposed as a promising technology for a sustainable energy source because water is abundant, sustainable, and carbon-free, and the electricity for the hydrogen production can be generated by wind and solar.[1, 2] However, the energy consumption in the hydrogen evolution reaction (HER, $2H^+ + 2e^− \rightarrow H_2$) is usually high due to its slow reaction kinetics, resulting in the need for efficient and durable electrocatalysts.[3] Platinum (Pt) is the most efficient electrocatalyst for HER, but its abundance is six orders of magnitude lower than aluminum, the most abundant metal,[4, 5] making it extremely expensive (ten million US$ per ton, Figure S1). As a result, the Pt or any other noble metal catalyst accounts for ~8% of the total stack cost of a proton exchange membrane electrolyzer.[6] Because of this reason, researchers have devoted great effort to exploring Pt-free or low-Pt electrocatalysts, including Pt single atoms,[7, 8] Pt-based alloys,[9, 10] metal carbides,[11-13] transition metal dichalcogenides (TMDCs),[14-20] metal phosphides,[6, 21] *etc*. For example, Liu *et al*. anchored Pt single atoms on carbon nanospheres, and the resulting material showed a comparable HER performance to commercial Pt/C but with less Pt needed.[7] King *et al.* loaded CoP nanoparticles on a high-surface-area carbon support, which showed an excellent activity and a long-term stability that were close to commercial Pt/C.[6] Although great

progress has been made in the low-Pt and Pt-free electrocatalysts, how to achieve the high-throughput production of such catalysts is still a challenge to using water electrolysis. Moreover, the current densities needed by industry are usually higher than 1000 mA cm$^{-2}$ and 500 mA cm$^{-2}$ for proton exchange membrane and alkaline electrolyzers,[13] requiring catalysts with good electrochemical, thermal, and mechanical stabilities, as well as abundant numbers of active sites.[6, 13, 22] These challenges have motivated the need for the high throughput production of efficient, durable, and cheap electrocatalysts for a high-current-density HER.

$MoS_2$ is promising for HER because of its high catalytic activity, good stability, and low price. Hinnemann *et al.* forecast that $MoS_2$ edges would be active for HER with a free adsorption energy of hydrogen close to zero,[23] which was then experimentally verified by Jaramillo *et al.* using two-dimensional (2D) $MoS_2$[15] because the 2D form had more exposed edges than the bulk material.[16, 24, 25] Our group has recently demonstrated that by combining surface chemistry and morphology engineering, $MoS_2$ showed a good HER performance at current density of 1000 mA cm$^{-2}$.[13] $MoS_2$ also has a good atmospheric thermal stability (up to 300 °C),[26] a good electrochemical stability under reducing potentials,[26] and is mechanically roubust,[27, 28] making it a promising catalyst for high-current density HER. The global proven reserve of Mo is about 500 times that of Pt and its price is three orders of magnitude lower (Figure S1), making the cheap production of $MoS_2$-based catalysts possible. $MoS_2$ naturally exists as the molybdenite mineral, with a global availability of 17,000,000 tons.[4] If one can use such a low-cost and abundant mineral to produce suitable catalysts, the overall cost of HER electrocatalysts would be significantly reduced. Despite its good performance and stability, most of the methods to produce $MoS_2$-based catalysts are energy intensive and/or difficult to scale-up due to the use of conditions such as high vacuum,[15] poisonous reactants,[24, 25] high pressure,[13, 26] or the poor high-current-density HER performance. Therefore, the production of efficient, cheap, and durable MoS$_2$-based catalysts by a high-throughput and scalable way is desired to make a real impact in HER electrochemical technology.

Here, we report a high-throughput scalable method for production of cheap yet high-performance MoS$_2$-based HER catalysts that work well at high current densities up to 1000 mA cm$^{-2}$. We first obtained 2D MoS$_2$ flakes by a scalable top-down exfoliation method, followed by a simple thermal treatment to prepare the catalyst, with both processes having the possibility of being scaled up for high-throughput production. The catalysts are 2D MoS$_2$ modified by Mo$_2$C nanoparticles on their edges and surfaces, and have a good HER performance with a high current density of 1000 mA cm$^{-2}$ at 454 mV, a small Tafel slope of 68 mV dec$^{-1}$, and good stability for 24 h. We also demonstrated the feasibility of the high-throughput production method by using a cheap molybdenite concentrate from a naturally existing earth-abundant mineral and found that the mineral catalysts also showed good HER performance at high current densities. The production rate of the electrocatalyst was as high as 1.3 g h$^{-1}$, one to two orders of magnitude higher than previous results (Tables S1), and the catalyst price was ~ 10 US\$ m$^{-2}$, around 30 times lower than than a commercial Pt/C electrocatalyst.

**Results**

**Preparation of MoS$_2$-based ink-type electrocatalysts**

A schematic of the process is shown in Figure 1a. The MoS$_2$-based catalyst was synthesized by a two-step method, (a) exfoliation of bulk MoS$_2$ into 2D flakes and (b) thermal treatment (see details in the "Methods" section). In brief, 2D MoS$_2$ was first prepared by exfoliating bulk MoS$_2$ (Figure S2) by an interMediate-Assisted Grinding Exfoliation (iMAGE) technique that was able to obtain 2D materials at the tonne scale.[28] Here we used a modified iMAGE technique that used $Mo_2C$ as the force intermediary to facilitate the exfoliation of $MoS_2$ because it has a low electrical resistivity (57 µΩ cm) and a high hardness (the Moh's hardness is 7).[29] The 2D $MoS_2$ flakes were then dispersed in water to obtain a catalyst dispersion, which is ink-type and is suitable to be integrated with robust industrial used techniques to produce large-area electrodes such as dip-coating, drop-casting, roll-to-roll printing, screen printing, and spray coating. Here, the exfoliated $MoS_2$ flakes together with $Mo_2C$ were loaded by dip coating onto supports with high surface areas (*e.g.*, carbon cloth, Ti foam, Cu foam) for the $CH_4/H_2$ thermal treatment, after which 2D $MoS_2$ became a suitable catalyst for the subsequent HER test (denoted as HC-$MoS_2$/$Mo_2C$, where 'HC' means $CH_4/H_2$ thermal treatment). To optimize the HER performance of the catalyst, the thermal treatment was divided into two parts: i) desulphurization using $H_2$ as the reacting gas to form S vacancies on 2D $MoS_2$, and ii) carburization using $CH_4/H_2$ as the reacting gases to produce $Mo_2C$ nanoparticles on 2D $MoS_2$.[13] In the first stage, S vacancies were formed by the reaction of $H_2$ with S in the $MoS_2$, removing S atoms to form $H_2S$, while in the second stage $CH_4$ reacted with Mo atoms near the vacancies to form $Mo_2C$ by dehydrogenation.[30] Both processes involved mild conditions and can be scaled-up for the high-throughput production of $MoS_2$-based catalysts. Examples of several $MoS_2$-based catalysts on different supports prepared by this method are shown in Figures 1b-d and S3.

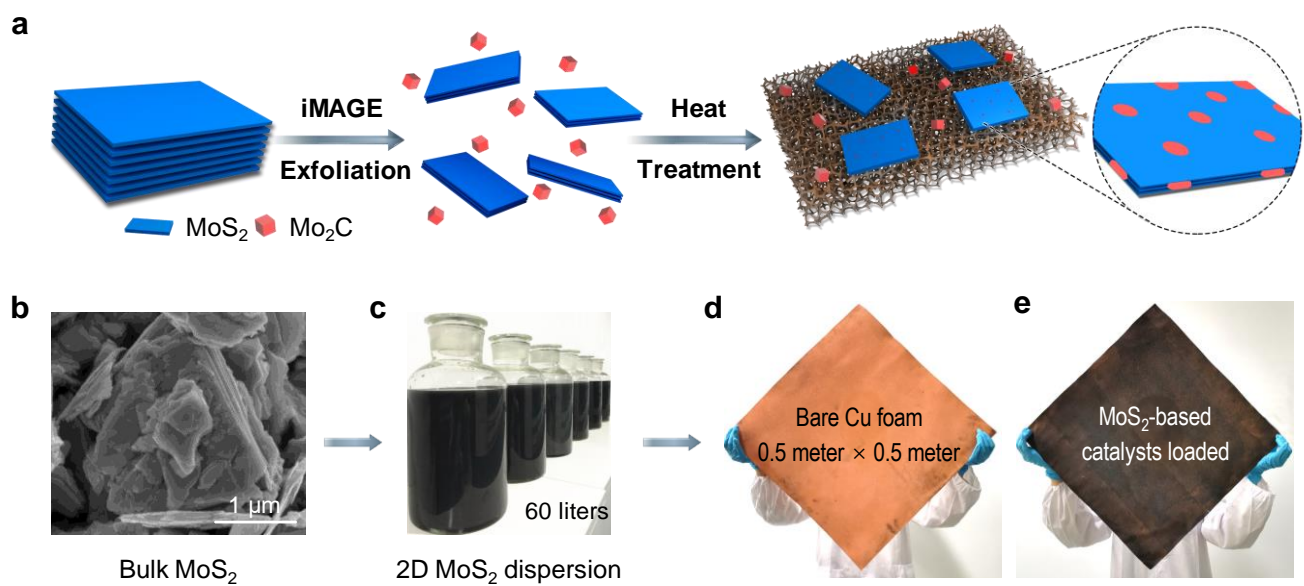

**Figure 1. High-throughput production of MoS₂-based ink-type electrocatalysts.** (a) Schematic of the fabrication method of MoS₂-based catalysts. (b) SEM image of the pristine MoS₂ powder, and photos of (c) 2D MoS₂ aqueous dispersion with a volume of 60 liters, (d) bare Cu foam and (e) MoS₂-based catalyst loaded on Cu foam.

**Structure characterization of the catalysts.**

We characterized the structures of the MoS₂ materials after the first exfoliation step and the second thermal treatment step. To expose more MoS₂ edges, we choose a bulk MoS₂ material with a relatively small lateral size (~1 μm). Atomic force microscopy (AFM) showed that after the first exfoliation step, the bulk MoS₂ was exfoliated into 2D MoS₂ flakes (Figure 2a) with an average thickness of 15 nm (Figure 2b) and an average lateral size of 0.6 μm (Figures 2c), consistent with the results of dynamic light scattering (DLS, Figure S4), transmission electron microscopy (TEM, Figure S5), and scanning electron microscopy (SEM, Figure S5). The uniformity of the 2D MoS₂ was demonstrated by combined AFM, SEM, and TEM characterization. In addition, results from high resolution transmission electron microscopy (HRTEM) and the corresponding fast Fourier transformation (FFT) showed the high quality of the 2D MoS$_2$ flakes without noticeable defects in their basal planes and edges (Figures 2d and S6). The above results show that in the first step the bulk MoS$_2$ was exfoliated into 2D MoS$_2$ flakes with good uniformity and high quality.

After the second thermal treatment step, we confirmed the formation of Mo$_2$C nanocrystals on the 2D MoS$_2$ by HRTEM, X-ray diffraction (XRD), and Raman spectroscopy. The HRTEM images show that in the final optimized HC-MoS$_2$/Mo$_2$C catalyst, Mo$_2$C nanocrystals were formed both on the basal planes and edges of 2D MoS$_2$ flakes (Figures 2e and S7). The growth of Mo$_2$C nanocrystals in these positions is easy to understand because S vacancies and unsaturated Mo existed there, which facilitated their formation. Higher magnification HRTEM images show typical lattice spacings of 0.29 nm and 0.24 nm, which respectively correspond to the (100) planes of MoS$_2$ (blue dotted box in Figure 2e) and the (002) planes of β-Mo$_2$C (red dotted box in Figure 2e). Statistical analysis shows that the Mo$_2$C nanocrystals (inset of Figure 2e) have an average diameter of 2.8 nm and a narrow size distribution, and are evenly distributed over the 2D MoS$_2$ basal planes. Regarding the XRD patterns (Figures 2f and S8), the MoS$_2$ (002) peak is quite sharp, indicating the high crystallinity of the MoS$_2$. The peaks at 34.5° and 38.1° correspond to the (100) and (002) planes of Mo$_2$C. Moreover, Raman spectroscopy was also used to identify the structures of the HC-MoS$_2$/Mo$_2$C (Figure 2g). In addition to the MoS$_2$ peaks, those at 812 and 987 cm$^{-1}$ indicate the formation of Mo$_2$C.[31] Taking these results together, we can say that the HC-MoS$_2$/Mo$_2$C is composed of 2D MoS$_2$ flakes with a large number of Mo$_2$C nanocrystals on their basal planes and edges, and good uniformity of the catalyst is achieved by this method.

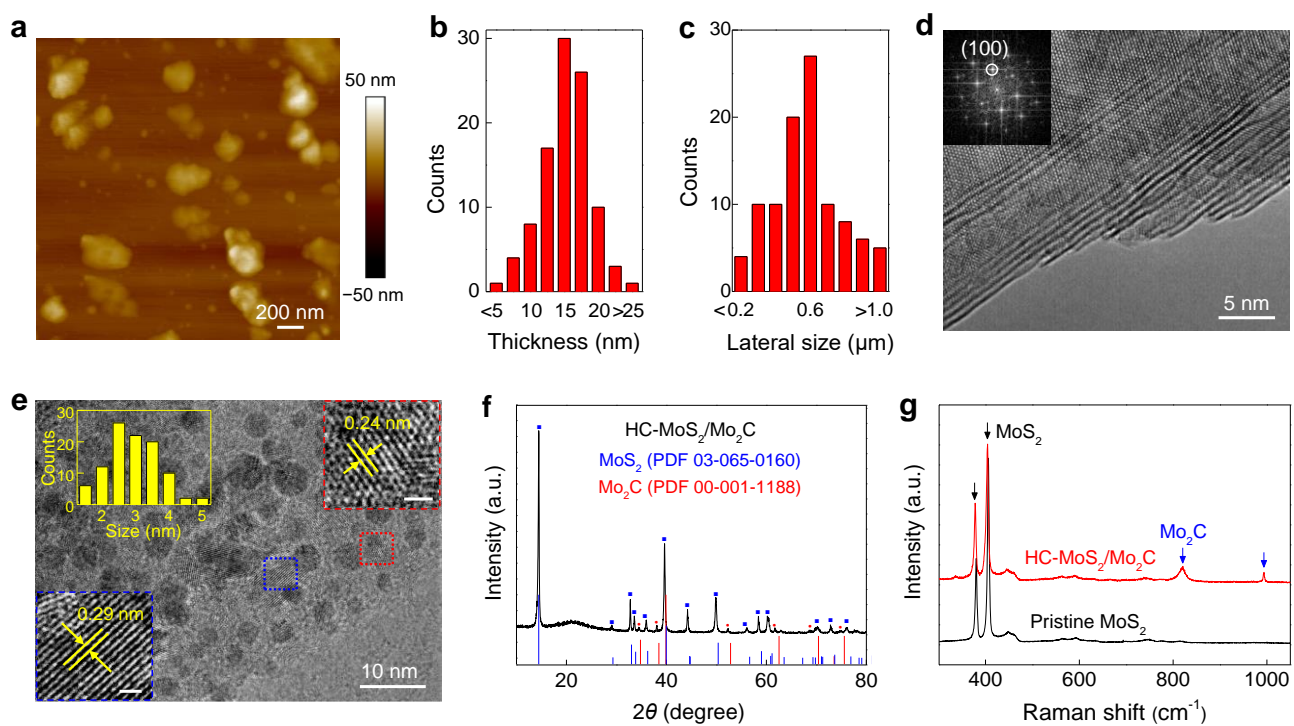

**Figure 2. Material characterization.** (a) AFM image and statistical analysis of (b) the lateral size and (c) the thickness of the 2D MoS$_2$ flakes. (d) HRTEM image of the 2D MoS$_2$. Inset is the corresponding fast Fourier transform (FFT) pattern. (e) HRTEM image of the HC-MoS$_2$/Mo$_2$C. The insets are a histogram of the lateral size of the Mo$_2$C nanocrystals, and two high magnification HRTEM images of a MoS$_2$ flake (the blue dotted box) and a Mo$_2$C nanocrystal (the red dotted box). The scale bars in the insets are 1 nm. (f) XRD pattern and (g) Raman spectra of the 2D MoS$_2$ and the HC-MoS$_2$/Mo$_2$C.

**HER performance at high current densities.**

To evaluate the electrochemical performance of the catalysts, we used a standard three-electrode electrolyzer. The distance between the working and reference electrodes was optimized to measure the HER performance at high current densities and the catalysts were tailored to ensure identical electrode surface areas (Figure S9). The HER performance of catalysts was compared by a parameter that combines the contributions from the overpotential ($\eta$) and the iR drop ($iR_s$) as in the following equation,

$$iR_s + \eta = E_{appl} - E_{eq}$$

where $E_{appl}$ and $E_{eq}$ are the practical applied and theoretical electrochemical equilibrium potentials (0 V *vs* RHE for HER). Several key parameters such as the nature of the supports, the loading amount of 2D MoS$_2$ flakes, as well as the temperature and time for the desulphurization and carburization processes have been systematically studied and optimized (Figures S3 and S10). Based on the best samples obtained, we characterized and compared the HER performance of the HC-MoS$_2$/Mo$_2$C with other three samples, a Pt foil, bare Cu foam, and 2D MoS$_2$ without the thermal treatment. Polarization curves of the four samples in a H$_2$SO$_4$ (0.5 M) solution are shown in Figure 3a and Movie S1. Compared to the landmark Pt, the HC-MoS$_2$/Mo$_2$C catalyst showed a higher onset potential and needed a larger $E_{appl} - E_{eq}$ to obtain an identical current density (*j*) when current density is smaller than 120 mA cm$^{-2}$. Tafel plots were used to study the rate-determining step of the catalysts (Figure 3b) and the results show that the HC-MoS$_2$/Mo$_2$C has a small slope of 68 mV dec$^{-1}$, indicating that the recombination of hydrogen is its rate-limiting step.[32] As current density increased to over 120 mA cm$^{-2}$, the HC-Mo$_2$S/Mo$_2$C required a much smaller $E_{appl} - E_{eq}$ to obtain an identical current density to the Pt. For instance, the HC-MoS$_2$/Mo$_2$C needed only 405 mV @ 500 mA cm$^{-2}$ and 454 mV @ 1000 mA cm$^{-2}$, whereas the respective values for Pt were 772 and 1191 mV. These results show the superior HER performance of the HC-MoS$_2$/Mo$_2$C at large current densities. Neither the 2D MoS$_2$ without the thermal treatment nor the bare Cu foam showed such a performance as the current density increased. Note that for the sample without the thermal treatment, the 2D MoS$_2$ flakes fell from the support as current density increased and made the catalytic performance even poorer, caused by the forces of H$_2$ bubbles (Figure S11 and Movie S2). In sharp contrast, the HC-MoS$_2$/Mo$_2$C was tightly attached to the Cu foam support and showed good mechanical stability at high current densities (Movie S1). An SEM image of the HC-MoS$_2$/Mo$_2$C showed a soldering-like phenomenon between the HC-MoS$_2$/Mo$_2$C and the Cu support, indicating that thermal treatment may be a good way to improve the mechanical robustness of electrocatalysts. In addition, the HC-MoS$_2$/Mo$_2$C catalyst showed a pH-universal HER activity and also worked well in a KOH (1.0 M) solution (Figure S12).

For practical applications, the high-current-density performance of catalysts is of vital importance. To evaluate the HER performance of the HC-MoS$_2$/Mo$_2$C in such conditions, we analyzed the relationships between current densities and $\Delta(E_{appl}-E_{eq})/\Delta\log|j|$, which can be viewed as a generalized slope and can be used to evaluate the performance of catalysts at high current densities.[13] The Pt showed an increasing $\Delta(E_{appl}-E_{eq})/\Delta\log|j|$ ratio as current density increased, whereas the HC-MoS$_2$/Mo$_2$C maintained a small ratio at different current densities, indicating that it had an excellent HER performance at high current densities (Figure 3c). To understand the different performances of the HC-MoS$_2$/Mo$_2$C and Pt electrocatalysts, their mass transfer abilities were studied. The diameters of the H$_2$ bubbles remained small on HC-MoS$_2$/Mo$_2$C (Figure 3d), while they became larger on Pt as current density increased.[13] Moreover, in HC-MoS$_2$/Mo$_2$C we did not observe the fluctuation phenomenon at high current densities that is seen for Pt (Figure 3a), which might be related to the fast removal of H$_2$ bubbles from the surface of HC-MoS$_2$/Mo$_2$C. We confirmed this point by studying the time-dependent (30 frames per second) transport of H$_2$ bubbles on HC-MoS$_2$/Mo$_2$C, which showed the departure diameters of bubbles on its surface remained small (Figure 3e). These results indicate the good mass transfer ability of HC-MoS$_2$/Mo$_2$C at high current densities. Besides mass transfer, we measured the electrochemically active surface areas (ECSA) of these catalysts, which showed that the value for HC-MoS$_2$/Mo$_2$C was higher than for the other three samples (Figure S13). Electrochemical impedance spectroscopy (EIS) showed that HC-MoS$_2$/Mo$_2$C had the lowest charge transfer resistance of the four samples (Figure S14). Compared with other reports, the HC-MoS$_2$/Mo$_2$C showed the best achievable HER performance at 1000 mA cm$^{-2}$ among all MoS$_2$-based and Pt-based catalysts (Figure 3f and Table S2). In addition, stability tests showed that HC-MoS$_2$/Mo$_2$C retained its HER performance over 24 h at 200 mA cm$^{-2}$ and also at 500 mA cm$^{-2}$ (Figure 3g), which was confirmed by SEM images (Figure S11). All these results show that HC-MoS$_2$/Mo$_2$C has good achievable HER performance, demonstrated by a small $E_{appl} - E_{eq}$, a small Tafel slope, and good stability at large current densities.

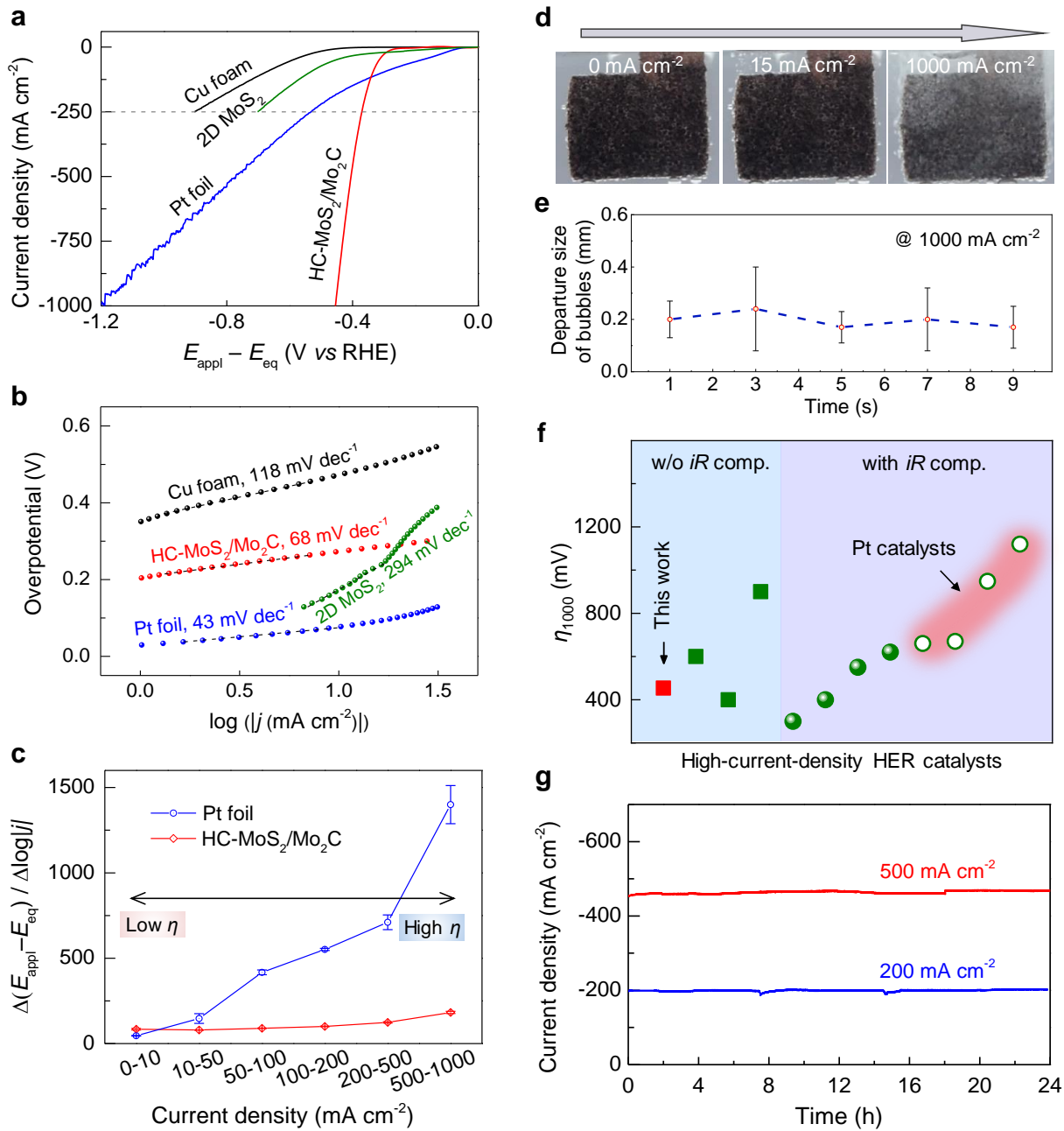

**Figure 3. Electrocatalytic performance of different catalysts for high-current-density HER.** (a) Polarization curves, (b) Tafel plots, and (c) $\Delta(E_{appl}-E_{eq})/\Delta\log|j|$ ratios of different catalysts in 0.5 M $H_2SO_4$ at a scan rate of 5 mV s$^{-1}$. (d) Photos and (e) corresponding size distribution of $H_2$ bubbles leaving the surfaces of HC-$MoS_2$/$Mo_2$C catalysts. (f) A comparison of the HER performance of HC-$MoS_2$/$Mo_2$C catalysts and those reported previously (see details in Table S2). The '(w/o) iR comp.' and 'with iR comp.' in (f) mean without (w/o) and with *iR* compensation, where the values are calculated by ($E_{appl}$–$E_{eq}$) and ($E_{appl}$–$E_{eq}$–$iR$). (g) Chronoamperometric response (i-t) curves for HER using HC-MoS$_2$/Mo$_2$C at current densities of 200 and 500 mA cm$^{-2}$ for 24 h.

**Production of cheap MoS$_2$ mineral catalysts for HER.**

To realize the large-scale use of electrolyzers, it is important to reduce the cost of all system components, and replacing expensive precious metals with cheap and durable catalysts is a critical step. The high-throughput production of such catalysts can pave the way for their cost reduction as well allowing their use on a large scale. To demonstrate the feasibility of high-throughput production method, we used molybdenite concentrate as the precursor, which is mainly composed of bulk MoS$_2$ in form of molybdenite mineral dug directly from an open-pit mine (Figure 4a). A molybdenite concentrate is generated by a preliminary floatation treatment that can be used to produce industrial-grade MoS$_2$ as well as high-purity MoS$_2$. Although this molybdenite concentrate still has a relatively low purity (Figures S15-S17), its low price gives it a huge potential for extensive use in the electrocatalyst industry. Its current price is only $10^{-2}$, $10^{-4}$, $10^{-5}$ times those of industrial-grade MoS$_2$, high-purity MoS$_2$, and Pt, respectively (Figure 4b). It is therefore reasonable to use this raw bulk material to explore industrial catalyst production. A thousand liters of 2D MoS$_2$ dispersions (~10 mg mL$^{-1}$) have now been produced in which the 2D MoS$_2$ flakes have an average size of 50-100 nm (Figure S18). We have therefore shown that 2D MoS$_2$ can be mass produced from a cheap molybdenite concentrate.

We then tested the HER performance of this MoS$_2$ mineral-based catalyst. Similar to the earlier experiment, a large amount of 2D MoS$_2$ (10 mg mL$^{-1}$) was loaded onto a conductive support followed by thermal treatment. Because this 2D MoS$_2$ was mass produced, the size of the working electrode can be made much bigger. We have therefore constructed working electrodes with areas of 1 and 5 cm$^2$, both of which showed good uniformity (Figure S19). To the best of our knowledge, this is the first time cheap HER catalysts have been fabricated from MoS$_2$ minerals. The difference in the HER performance between the MoS$_2$ mineral-based and high-purity MoS$_2$-based catalysts has been measured. The MoS$_2$ mineral-based catalyst showed a good HER performance at high current densities. For example, it showed a 554 mV @ 1000 mA cm$^{-2}$, only 22% below the value for HC-MoS$_2$/Mo$_2$C (Figure S20). The good HER performance of the bigger electrode was further demonstrated by EIS (Figure S21), which is quite encouraging because a large amount of HER-inert impurities, *e.g.*, silicates, are present in MoS$_2$ concentrates and the catalytic performance can be further improved as discussed in Figure S10. The polarization curves of different working electrodes in a H$_2$SO$_4$ (0.5 M) solution are shown in Figure 4c. They show that the 5 cm$^2$ working electrode achieves a current density of 1000 mA cm$^{-2}$ at 390 mV, which is 184 mV less than for a 1 cm$^2$ electrode. We also calculated the H$_2$ production rate for different working electrodes (Figures 4d and Table S3) and found that the performance of the 5 cm$^2$ electrode (38 mg h$^{-1}$) was an almost perfect scale-up from the 1 cm$^2$ electrode (7.4 mg h$^{-1}$), indicating the possibility of industrial use. Furthermore, the MoS$_2$ mineral-based catalyst exhibited good durability, maintaining its performance at a current density of 500 mA cm$^{-2}$ for more than 24 h (Figure 4e).

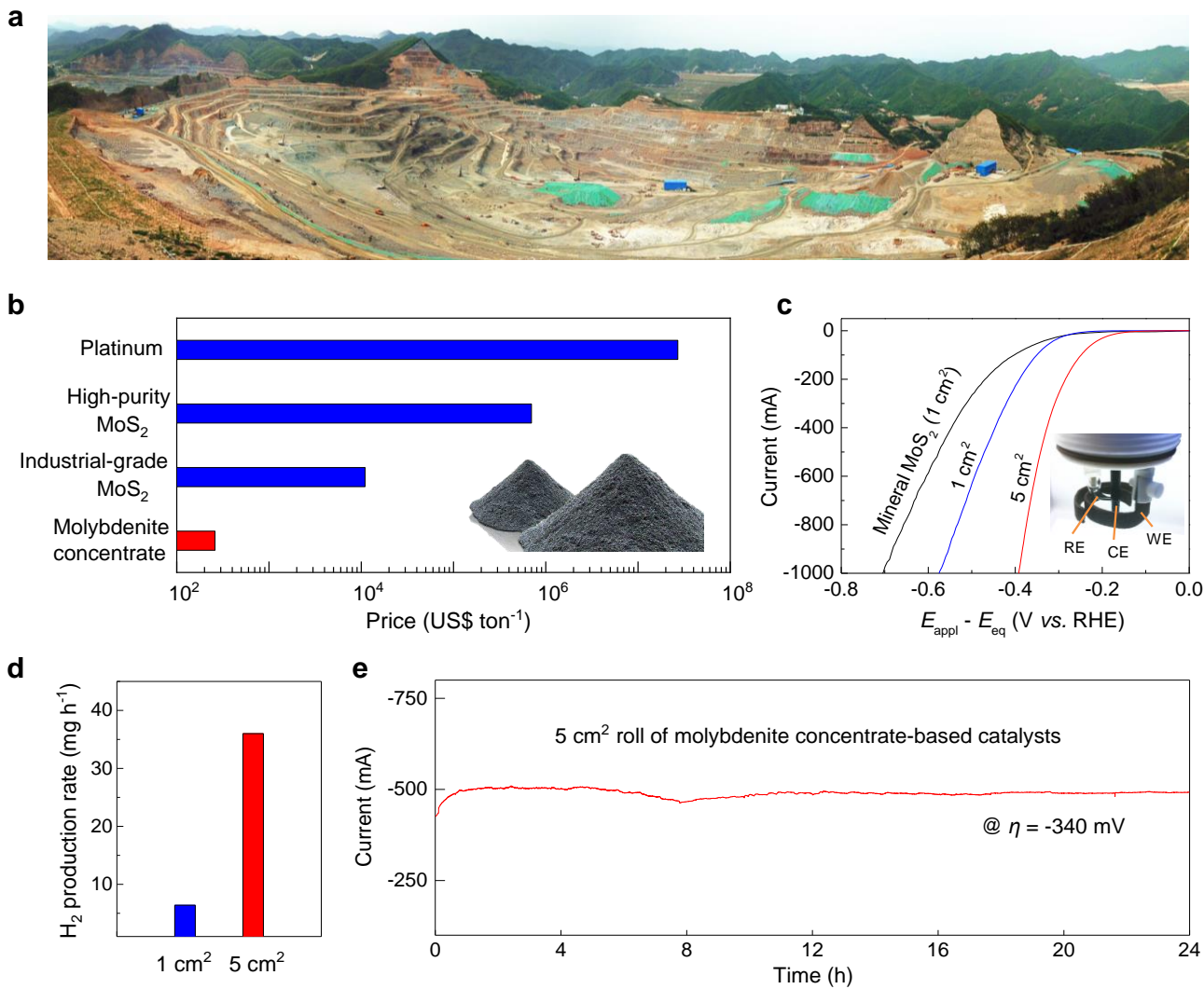

**Figure 4. MoS₂ mineral catalysts from cheap molybdenite concentrates for high-current-density HER.** (a) Bird's eye view of the Sandaozhuang open-pit molybdenite mine in Luoyang, China. Reproduced from [28] with permission from the Oxford University Press, copyright 2019. (b) Commodity price differences between platinum, high-purity MoS₂, industrial-grade MoS₂, and molybdenite concentrate. (c) Polarization curves of the different catalysts in 0.5 M H₂SO₄ at a scan rate of 5 mV s$^{-1}$. Inset is an optical image of a three-electrode electrolyzer for larger working electrodes. (d) H₂ production rate of different molybdenite concentrate-based and high purity MoS₂-based catalysts to produce hydrogen under -390 mV of $E_{appl}-E_{eq}$. (e) Chronoamperometric response (i-t) curve of a 5 cm² roll of cheap mineral-based catalyst for HER under -340 mV of $E_{appl.}-E_{eq.}$ over 24 h.

The fabrication method for the HC-MoS$_2$/Mo$_2$C catalyst has noticeable advantages over other methods such as solvothermal synthesis, gas-solid reaction, tip sonication, and Li intercalation (Figure 5a and Table S1). For example, the production rate of exfoliated 2D MoS$_2$ is ~1.3 g h$^{-1}$, which is one to two orders of magnitude higher than those of other methods for synthesizing MoS$_2$-based or even other TMDC-based catalysts using the most ideal assumptions. Our catalyst also works well at high current densities up to 1000 mA cm$^{-2}$ giving it a big advantage in both production rate and working current density. The 2D MoS$_2$ ink also has the advantage of being able to be added to electrodes by spraying and dipping. The MoS$_2$ mineral-based catalyst also has a low price of only 10 US\$ m$^{-2}$ excluding the cost of the support, almost 30 times lower than commercial Pt/C catalysts (Figure 5b and Table S4). Possible replacement of the Mo$_2$C additive by cheaper materials would further reduce the overall cost. While the result presented here is a notable achievement, we note that a more comprehensive analysis will be the subject for future study. Taken together, these results indicate the huge potential of our high-throughput method for fabricating cheap, high-performance, and durable MoS$_2$-based catalysts from minerals that are suitable for large-scale H$_2$ production.

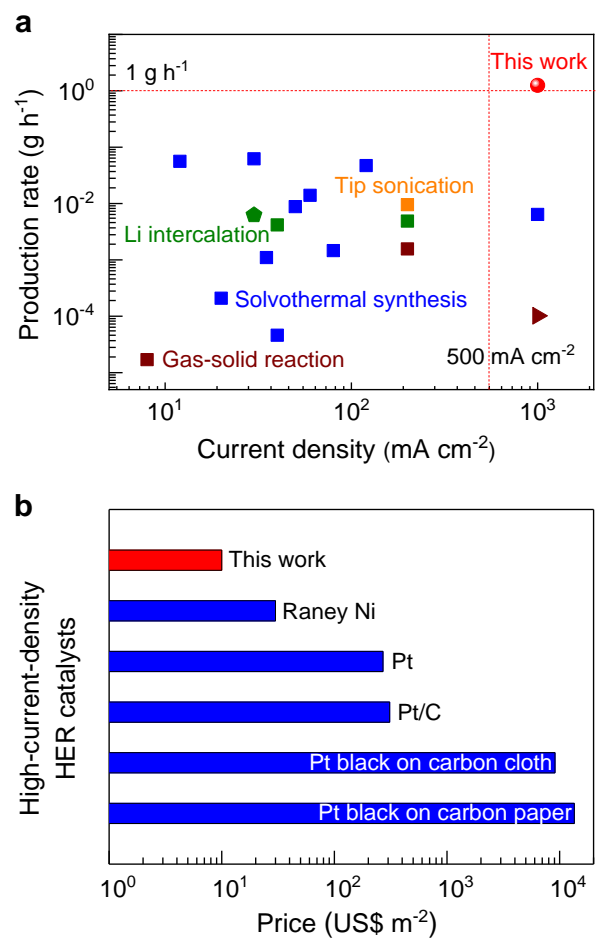

**Figure 5. High-efficiency and low-cost production of MoS$_2$ mineral-based catalysts from cheap molybdenite concentrate for high-current-density HER.** (a) A comparison of the fabrication rate and highest tested current density of transition metal dichalcogenide-based HER catalysts by different methods. Details for data points are shown in Table S1. (b) A comparison of the catalyst cost compared to commercial electrocatalysts for HER, showing the ultralow cost of the HC-MoS$_2$/Mo$_2$C catalyst. Note that the support costs of HC-MoS$_2$/Mo$_2$C, Raney Ni, Pt and Pt/C catalysts are excluded.

**Discussion**

We have reported a high-throughput and scalable method for production of MoS$_2$-based ink-type electrocatalysts, which combines scalable top-down exfoliation and simple thermal treatment. The catalysts exhibit decent performance for high-current-density HER, which is also verified by using cheap and abundant MoS$_2$ mineral. This is the first time a catalyst with good high-current-density HER performance has been produced from cheap molybdenite minerals, giving it a huge potential for large-scale industry hydrogen production. Besides molybdenite minerals for HER reported in this work, the method could be extended to the exfoliation of other layer materials from abundant natural resources for the mass production of electrocatalysts toward different electrochemical technologies.

**Methods**

**Exfoliation of 2D MoS$_2$.** All chemicals were used as received without further purification. The 2D MoS$_2$ was exfoliated by modifying a method previously reported.[28] The bulk MoS$_2$ (10 g, with an average particle size of 1-2 μm) and the Mo$_2$C (20 g, with an average particle size of 44 μm) powders were mixed together and added into a rotational grinding apparatus (Retsch RM 200, Germany). As a result the MoS$_2$ was exfoliated into 2D MoS$_2$ after grinding for 9 h in an the ambient atmosphere. For mineral exfoliation, molybdenite concentrate (20 g, from the Sandaozhuang open-pit mine, Luoyang, China) was used in place of the commercial MoS$_2$ powder.

**Thermal treatment of 2D MoS$_2$ to form HC-MoS$_2$/Mo$_2$C electrocatalyst.** First, the 2D MoS$_2$ was loaded onto a Cu foam that had been cleaned by diluted HCl, water, and acetone. The exfoliated 2D MoS$_2$ (600 mg) was added to a mixture of ethanol (18 mL) and water (2 mL) and shaken for 10 s to produce a suspension was dropped onto the Cu foam (1 × 1 cm$^2$) with different mass loadings (2−12 mg 2D MoS$_2$). Other conductive supports, Ti foam and carbon cloth, were also examined. Second, the conductive supports loaded with 2D MoS$_2$ were heat treated. In a typical procedure, Cu foam loaded with 2D MoS$_2$ was placed in a quartz boat in the center of a 1.5 in. diameter quartz tube furnace. For H$_2$ treatment, the furnace was first kept at given temperature (650, 750, or 850 °C) in a mixture of Ar (25 sccm) and H$_2$ (7.5 sccm) for different times (10, 30, 100, or 180 min), after which S vacancies had been formed in the 2D MoS$_2$. This was followed by CH$_4$ treatment, in which the H$_2$-treated sample was held at an optimum temperature of 750 °C in a mixture of Ar (25 sccm), H$_2$ (2.5 sccm) and CH$_4$ (2.5 sccm) for different times (10, 30, 60 or 180 min) in order to partially convert the MoS$_2$ into Mo$_2$C nanocrystals. The final optimized sample is denoted HC-MoS$_2$/Mo$_2$C.

**Materials characterization.** The surface morphology of the HC-MoS$_2$/Mo$_2$C samples was characterized by SEM (5 kV, Hitachi SU8010, Japan). The thickness of the 2D MoS$_2$ flakes was measured by AFM (Bruker Dimension Icon, Germany). TEM and HRTEM were carried out by using an electron acceleration voltage of 300 kV (FEI Tecnai F30, USA). Structural and chemical analyses of the samples were performed by powder XRD (Cu Kα radiation, λ = 0.15418 nm, Bruker D8 Advance, Germany), while Raman spectra were collected using 532 nm laser as the excitation light with a beam size of ~1 μm (Horiba LabRAB HR800, Japan).

**Electrochemical measurements.** A standard three-electrode electrolyzer with H$_2$SO$_4$ (0.5 M) or KOH (1.0 M) was used in all tests, with a saturated calomel electrode (SCE) and a graphite rod as the reference and counter electrodes, respectively. Pt counter electrode is used for taking videos. The scan rates were 5 mV s$^{-1}$ for the linear sweep voltammetry tests and the scan rates for the cyclic voltammetry tests have been noted in the captions of the related figures.

**Acknowledgement**

We acknowledge support from the National Natural Science Foundation of China (Nos. 51722206 and 51920105002), the Youth 1000-Talent Program of China, Guangdong Innovative and Entrepreneurial Research Team Program (No. 2017ZT07C341), the Bureau of Industry and Information Technology of Shenzhen for the "2017 Graphene Manufacturing Innovation Center Project" (No. 201901171523), and the Development and Reform Commission of Shenzhen Municipality for the development of the "Low-Dimensional Materials and Devices" discipline.

**Author contributions**

C.Z., Y.L., H.C., and B.L. conceived the idea. C.Z. synthesized materials and performed XRD, Raman, SEM, and AFM characterization and electrochemical tests. L.Y. helped in materials exfoliation. J.T. performed TEM characterization and analysis. Y.L., F.Y., and Z.Z. took part in the electrochemical measurements and discussion. B.L. supervised the project and directed the research. C.Z., Y.L., H.C. and B.L. interpreted the results. C.Z., Y.L., and B.L. and wrote the manuscript with feedbacks from the other authors.

**Additional information**

**Supplementary Information** accompanies this paper at http://www.nature.com/nature

communications.

**Competing financial interests:** Patents related to this research have been filed by Tsinghua-Berkeley Shenzhen Institute, Tsinghua University. The University's policy is to share financial rewards from the exploitation of patents with the inventors.